\documentclass[11pt]{article}
\usepackage{amsmath,amssymb,epsf}
\usepackage{setspace}
\usepackage[compress]{cite}

\newcommand{\be}{\begin{equation}}
\newcommand{\ee}{\end{equation}}

\newcommand*\samethanks[1][\value{footnote}]{\footnotemark[#1]}
\begin{document}
\setstretch{1.15}
\title{\bf\large A computationally-efficient sandbox algorithm for multifractal analysis of large-scale complex networks with tens of millions of nodes}

\author{Yuemin Ding$^{1, 2}$\thanks{Joint first authors: Y. Ding and J.-L. Liu}, Jin-Long Liu$^{3}$\samethanks, Xiaohui Li$^{4, 2}$, Yu-Chu Tian$^{2}$\thanks{Joint corresponding authors: Y.-C. Tian (email: y.tian@qut.edu.au) and Z.-G. Yu (email: yuzuguo@aliyun.com)}, and Zu-Guo Yu$^{2, 3}$\samethanks\\
	{\it\small $^{1}$School of Computer Science and Engineering, Tianjin University of Technology,}\\
	{\it\small Tianjin 300384, China.}\\
	{\it\small $^{2}$School of Computer Science, Queensland University of Technology, Brisbane QLD 4000, Australia.}\\
	{\it\small $^{3}$Key Laboratory of Intelligent Computing and Information Processing of Ministry of Education and}\\
	{\it\small Hunan Key Laboratory for Computation and Simulation in Science and Engineering,}\\
	{\it\small Xiangtan University, Xiangtan, Hunan 411105, China.}\\
	{\it\small $^{4}$School of Information Science and Engineering, Wuhan University of Science and Technology,}\\
	{\it\small Wuhan, Hubei 430081, China.}
}
\date{}
\maketitle

\begin{abstract}
The fractality of complex networks has attracted much attention with extensive investigations over the last $15$ years. As a generalization of fractal analysis, multifractal analysis (MFA) is a useful tool to systematically describe the spatial heterogeneity of both theoretical and experimental fractal patterns. One of the widely used methods for fractal analysis is box-covering. It uses the minimum number of covering boxes to calculate the fractal dimension of complex networks, and is known to be NP-hard. More severely, in comparison with fractal analysis algorithms, MFA algorithms have much higher computational complexity. Among various MFA algorithms for complex networks, the sandbox MFA algorithm behaves with the best computational efficiency. However, the existing sandbox algorithm is still computationally expensive. Thus, so far it has only been applied to small-scale complex networks of the size of about tens of thousands of nodes. It becomes challenging to implement the MFA for large-scale networks with tens of millions of nodes. It is also not clear whether or not MFA results can be improved by a largely increased size of a theoretical network. To tackle these challenges, a computationally-efficient sandbox algorithm (CESA) is presented in this paper for MFA of large-scale networks. Distinct from the existing sandbox algorithm that uses the shortest-path distance matrix to obtain the required information for MFA of complex networks, our CESA employs the breadth-first search (BFS) technique to directly search the neighbor nodes of each layer of center nodes, and then to retrieve the required information. Our CESA's input is a sparse data structure derived from the compressed sparse row (CSR) format designed for compressed storage of the adjacency matrix of large-scale network. A theoretical analysis reveals that the CESA reduces the time complexity of the existing sandbox algorithm from cubic to quadratic, and also improves the space complexity from quadratic to linear. MFA experiments are performed for typical complex networks to verify our CESA. The CESA is demonstrated to be effective, efficient and feasible through the MFA results of ($u$,$v$)-flower model networks from the 5th to the 12th generations. It enables us to study the multifractality of networks of the size of about 11 million nodes with a normal desktop computer. Furthermore, we have also found that increasing the size of ($u$,$v$)-flower model network does improve the accuracy of MFA results. Finally, our CESA is applied to a few typical real-world networks of large scale.


\end{abstract}

%%%%%%%%%%%%%%%%%%%%%%%%%%%%%%%%%%====================================================%%%%%%%%%%%%%%%%%%%%%%%%%%%%%%%
\clearpage\newpage
\setstretch{1.3}
\section{Introduction}
\label{Sec: Introduction}

Since Song \textit{et al.} \cite{SHM05} revealed the existence of the self-similarity in complex networks, the fractality of complex networks has attracted much attention with extensive investigations. This is due to its potential applications in various areas of science and technology \cite{SHM06, GSHM07, CHKSS07, RHB07, RSM10, GMS12, LYZ14, LYA14, CKHFM17, WWXCWBYZ17, WWXHLZ19,  QZYL19}. For some complex systems with an inhomogeneous distribution of local density of their certain physical quantities, however, a unique fractal dimension is not sufficient to characterize their complexity. As a generalization of fractal analysis, multifractal analysis (MFA) is a useful and more powerful tool to systematically describe the spatial heterogeneity of both theoretical and experimental fractal objects in many fields, such as economic systems \cite{Anh00, OEHJWSK12}, biological systems \cite{IAGHRSS99, YAL01, YAL03}, and physics and chemistry \cite{SM88, YAE09, LBR19}. In recent years, some studies have also focused on the MFA of complex networks. MFA has been shown to have better performance than fractal analysis in characterizing the complexity of model and real-world networks \cite{LYZ14, PLV10, FY11, WYA12, MMAB15, LYA15, SLYL15, MHKPEREGO16, XB17, HYA17, LWYX17, PRRC18, MMMHS18, LYA19, VMR19, PM20}. Thus, if a network possesses the multifractal property, we can use the generalized fractal dimensions $D_{q}$, instead of a single fractal dimension $D_{0}$, to unfold effectively the self-similar structure of the network, thus capturing the fluctuations of local node density in the network.

A few MFA algorithms have been proposed so far to calculate the generalized fractal dimensions $D_{q}$ or mass exponents $\tau_{q}$ of complex networks \cite{LYZ14, FY11, WYA12, LYA15, SLYL15, LWYX17}. The most widely used ones include the compact-box-burning (CBB) algorithm \cite{FY11}, the improved box-counting (IBC) algorithm \cite{LYZ14}, and the modified sandbox algorithm \cite{LYA15}. Box-covering with minimum number of boxes for calculating the fractal dimension of complex networks is known to be an NP-hard problem. More severely, in comparison with fractal analysis algorithms, MFA algorithms have much higher computational complexity, making MFA challenging. As described in \cite{LYA15}, the CBB and IBC algorithms must take a large amount of CPU time and memory resources to find the minimum possible number of boxes for covering the entire network because finding the minimal box-covering of a network is known to be NP-hard. Among various MFA algorithms, the existing sandbox algorithm behaves with the best computational efficiency for MFA of small-scale networks as experimentally illustrated in Ref. \cite{LYA15}. This is because that the sandbox algorithm only randomly selects a number of nodes on a network as the center nodes of sandboxes and then counts the number of nodes in each sandbox within a given radius for MFA. Therefore, the existing sandbox algorithm and its improved versions have been widely used to the calculation of the mass exponents $\tau_{q}$ or the generalized fractal dimensions $D_{q}$ of different types of complex networks \cite{CKHFM17, LYA15, SLYL15, MHKPEREGO16, HYA17, LWYX17, PRRC18, MMMHS18, LYA19}. The calculated results are then used for the investigation into the fractal and multifractal properties of the networks.

Despite of its advantages, the existing sandbox algorithm is still computationally expensive for large-scale complex networks. So far, it has only been applied to small-scale networks. For example, the multifractality of the 5th generation minimal model network with $12,501$ nodes has been studied by using the existing sandbox algorithm in Ref. \cite{LYA15}. Song \textit{et al.} \cite{SLYL15} have proposed an improved sandbox algorithm to uncover the multifractal property of the weighted Astrophysics collaboration network with $16,706$ nodes. Huang \textit{et al.} \cite{HYA17} have applied the improved sandbox algorithm to the MFA of the 7th generation weighted Koch networks with $32,769$ nodes. Overall, the largest size of complex networks reported using the existing sandbox algorithm and its improved versions for MFA is in the order of tens of thousands nodes. As will be seen later in Section \ref{Sec: Existing-SA}, for a complex network with $N$ nodes, the time complexity and space complexity of the existing sandbox algorithm are $O(N^{3})$ and $O(N^{2})$, respectively. With the increase in the network size $N$, the required computing resources characterized by $O(N^{2})$ increases rapidly. For example, for a complex network with a million (i.e., $10^6$) nodes, the required memory space resource is in the order of a few terabytes ($\approx 3.6$ TB), which is not available in normal desktop computers or even some high-performance computers. Thus, it becomes challenging to conduct the MFA for large-scale complex networks with millions of nodes or even tens of millions of nodes, such as the social networks, road networks, and autonomous systems graphs provided on Stanford Large Network Dataset Collection \cite{SLNDC}. Moreover, it is also not clear whether or not MFA results can be improved by an increased size $N$ of theoretical networks. All of these require a computationally-efficient algorithm to conduct the MFA for large-scale complex networks experimentally.

To tackle these challenges, a computationally-efficient sandbox algorithm (CESA) is proposed in this paper for MFA of large-scale complex networks. We first briefly recapitulate the existing sandbox algorithm for MFA of complex networks, developing some insights into its time and space complexities in the Section \ref{Sec: Existing-SA}. Then, Section \ref{Sec: CESA} presents the CESA with a theoretical analysis of its time and space complexities. This is followed by Section \ref{Sec: Experiment} on some MFA experiments on a normal desktop computer with a 4-core CPU and 16 GB memory. The experiments are presented to verify the CESA, to evaluate the impact of network size on accuracy of the algorithm, and to investigate the computational performance with verification networks generated from the ($u$,$v$)-flower network model. The CESA is also applied to a few real-world complex networks of large scale. Finally, Section \ref{Sec: Conclusions} concludes the paper.

%%%%%%%%%%%%%%%%%%%%%%%%%%%%%%%%%%====================================================%%%%%%%%%%%%%%%%%%%%%%%%%%%%%%%
\section{Insights into the complexity of the existing sandbox algorithm}
\label{Sec: Existing-SA}

This section briefly reviews the existing sandbox algorithm \cite{LYA15} and provides some insights into its complexity. Meanwhile, we also analyze the main factors that lead to huge computational burden for the existing sandbox algorithm to perform the MFA of large-scale complex networks.

%-------------------------------------------------------------------------------------------------------------------%
\subsection{The existing sandbox algorithm}
\label{Subsec: Sandbox}

According to the existing sandbox algorithm \cite{LYA15}, the generalized fractal dimensions $D_{q}$ of a complex network $G$ are defined as
\begin{equation}\label{Eq: MFA-Dq}
  D_q =\lim\limits_{r \to 0} \frac{\ln\langle[M_i(r)/N]^{q-1}\rangle}{\ln(r/d)} \frac{1}{q-1},~~q \in \Re,~~q\neq1,
\end{equation}
where $d$ denotes the diameter of the network $G$, $M_i(r)$ is the number of nodes covered by the sandbox with center node $i$ and radius $r~(1\leq r\leq d)$. $M_i(r)$ is one of the key parameters for estimating the generalized fractal dimensions $D_{q}$ of the network $G$. The pair of angle brackets $\langle \cdot \rangle$ denotes the operation of taking a statistical average over randomly chosen centers of the sandboxes. As shown in Eq. (\ref{Eq: MFA-Dq}), $D_{q}$ as a set of various dimensions describes the distribution of the measures of these sandboxes. It reflects the fluctuations of local node density in the network $G$. In particular, $D_0$, $D_1$, and $D_2$ represent the fractal dimension (or box-counting dimension), information dimension, and correlation dimension, respectively. In Eq. (\ref{Eq: MFA-Dq}), the information dimension $D_1$ cannot be directly calculated because $q \neq 1$. In practice, the generalized fractal dimensions $D_q~(q \rightarrow 1)$ are firstly calculated. After that, the interpolation method is used to obtain $D_1$. As a matter of fact, we usually rewrite Eq. (\ref{Eq: MFA-Dq}) as
\begin{equation}\label{Eq: MFA-LR}
    \ln(\langle[M_i(r)]^{q-1}\rangle) \propto D_{q}(q-1)\ln(r/d)+(q-1)\ln(N).
\end{equation}
If the network $G$ takes a multifractal structure, we can estimate its mass exponents $\tau_{q}$ numerically through a linear regression of $\ln(\langle[M_i(r)]^{q-1}\rangle)$ against $\ln(r/d)$, and calculate its generalized fractal dimensions $D_{q}$ through a linear regression of $\ln(\langle[M_i(r)]^{q-1}\rangle)/(q-1)$ against $\ln(r/d)$ \cite{LYA15}, respectively. Of course, we can also obtain the generalized fractal dimensions $D_{q}$ according to $\tau_{q}$ and $D_{q}=\tau_{q}/(q-1)$ for $q \neq 1$.

The existing sandbox algorithm for the MFA of network $G$ requires an input that is the shortest-path distance matrix of the network. Therefore, it is essential to calculate this shortest-path distance matrix by using some algorithms, e.g., the Floyd's algorithm \cite{Floyd62}. Thus, the main steps of the existing sandbox algorithm are described as follows:
\begin{enumerate}
\item[(i)] Calculate the shortest-path distance matrix of network $G$;

\item[(ii)] Set the radius $r~(1\leq r\leq d)$ of the sandbox;

\item[(iii)] A number of nodes are randomly selected as the centers of sandboxes to form the center-node set $S_{c}(r)$;

\item[(iv)] For each center node $i$ ($i \in S_{c}(r)$), count the number of nodes $M_i(r)$ covered by the sandbox with center node $i$ and radius $r$ according to the shortest-path distance matrix of network $G$;

\item[(v)] For each $q$, calculate the statistical average $\langle [M_{i}(r)]^{q-1} \rangle$ of $[M_{i}(r)]^{q-1}$ over all sandboxes of radius $r$;

\item[(vi)] For all different values of $r$, repeat steps (iii) to (v) to calculate the statistical averages $\langle [M_{i}(r)]^{q-1} \rangle$;

\item[(vii)] Calculate $\tau_{q}$ or $D_{q}$ with a linear regression according to Eq. (\ref{Eq: MFA-LR}).
\end{enumerate}

\begin{figure}[tbp]
\centerline{\epsfxsize=11cm \epsfbox{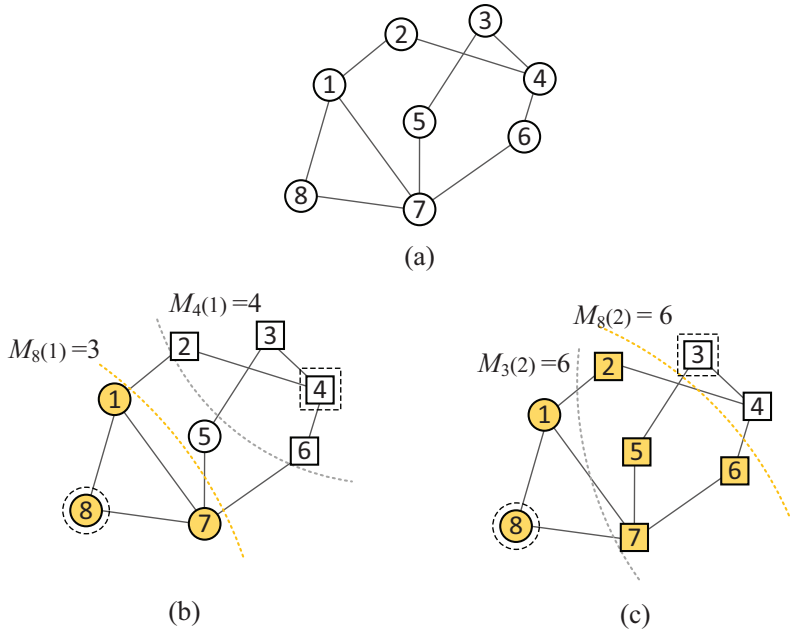}}
\caption{An example of the existing sandbox scheme.}
\label{Fig: Sandbox}
\end{figure}

This process is illustrated with a simple example in Fig. \ref{Fig: Sandbox}. For a simple network given in Fig. \ref{Fig: Sandbox}(a), assume that nodes $4$ and $8$ are selected as the center nodes when $r=1$. In this case, the number of nodes in the sandbox of center node $4$, $M_4(1)$, is $4$, and similarly, for center node $8$, $M_8(1)=3$, as shown in Fig. \ref{Fig: Sandbox}(b). Furthermore, assume that nodes $3$ and $8$ are chosen as the center nodes when $r=2$, $M_8(2)=6$ and $M_3(2)=6$ can be both determined as given in Fig. \ref{Fig: Sandbox}(c).

%-------------------------------------------------------------------------------------------------------------------%
\subsection{Complexity analysis of the existing sandbox algorithm}

To demonstrate the computational burden of the existing sandbox algorithm reviewed above, its computational complexity is analyzed. As usual, we assume that the parameter $q$ is in the range of [-10,10] with a step of $1/3$ in this study. Let $N_q$ denote the number of values of $q$ and $N_c$ denote the number of center nodes in the set $S_{c}(r)$. The sandbox algorithm takes the shortest-path distance matrix as its input. For a network $G$ of size $N$, it is known that the time complexity of the computation for the shortest-path distance matrix is $O(N^3)$ and the space complexity is $O(N^2)$. The time complexity of setting the value of $r$ from 1 to $d$ in the steps (ii) through (vi) is $O(d)$. The time complexity of step (iv) is $O(N_{c}N)$. The time complexity of step (v) is $O(N_{q}N_{c})$. Therefore, the overall time complexity of the algorithm can be expressed as $O(N^3+d(N_cN+N_qN_c))$. Here, $d$ indicates the network diameter. It is usually much smaller than the total number of nodes $N$ of network $G$. $N_q$ is constant and does not increase with the network size $N$. $N_c$ is usually set to be proportional to the network size $N$. Therefore, the overall time complexity of the algorithm can be further expressed as $O(N^3)$. The overall space complexity of the algorithm is $O(N^2)$, which is mainly determined by the size of the shortest-path distance matrix of network $G$. This implies that the required CPU time and memory space for the algorithm increases rapidly with the increase in the network size $N$. Consequently, the computing and memory burden is considerably heavy for large-scale complex networks. Because the shortest-path distance matrix of network $G$ is a full matrix, it cannot be compressed easily and is thus memory consuming. For an unweighed network $G$ with $100,000$ nodes as the example, the required memory space for its shortest-path distance matrix is in the order of a few tens of gigabytes. Our experimental tests show that the actual memory requirement for this network is approximately $37.3$ GB, challenging normal desktop computers. This is also the main reason why the existing sandbox algorithm has only been applied to small-scale networks so far.

%%%%%%%%%%%%%%%%%%%%%%%%%%%%%%%%%%====================================================%%%%%%%%%%%%%%%%%%%%%%%%%%%%%%%
\section{Computationally-efficient sandbox algorithm}
\label{Sec: CESA}

As shown in Eq. (\ref{Eq: MFA-LR}), the $M_i(r)$, namely the number of nodes covered by the sandbox with center node $i$ and radius $r$, is a key parameter for calculating the mass exponents $\tau_q$ of network $G$. It is directly obtained according to the shortest-path distance matrix of network $G$ as described in step (iv) of the existing sandbox algorithm. In fact, the $M_i(r)$ can also be calculated by accumulating the number of neighbor nodes from the $0$th layer to the $r$th layer of center node $i$. Here, the $l$th layer neighbor nodes of node $i$ are these nodes whose distance from the node $i$ equals to $l$. It is known that the breadth-first search (BFS) is an algorithm for searching tree or graph data structures \cite{CLRS09}. It starts from a root node, and then searches its neighbor nodes before searching the next layer neighbors. Therefore, we can apply the BFS algorithm with the center node $i$ as the root to obtain the neighbor nodes of each layer of center node $i$. Thus, the key parameter $M_i(r)$ can be easily calculated. As will be seen later in Subsection \ref{Subsec: Algorithm design}, another parameter $d$ does not affect the MFA results. And it can also be approximately estimated through applying BFS algorithm to all center nodes in the center-node set $S_c$. Therefore, we can redesign the computational process of sandbox scheme by directly searching the neighbor nodes of each layer of center node $i$ ($i \in S_c$). The new computational scheme eliminates the process of getting the shortest-path distance matrix, thus reducing the computational complexity of the existing sandbox algorithm.

With the consideration that typical complex networks are sparse networks, the usage of sparse matrix as the representation of network $G$ has the potential to significantly reduce the space complexity. The compressed sparse row (CSR) format, which is the current standard storage format for sparse matrices in numerical analysis and computer science, can reduce the substantial memory requirement, and enable fast row access \cite{Saad03}. It is convenient to extract the elements in each row. This is also beneficial for the design of the new computational scheme. Therefore, the CSR format of the adjacency matrix of network $G$ can be used as the input of the new computational scheme to reduce the space complexity. We call the new computational scheme a computationally-efficient sandbox algorithm (CESA).

%-------------------------------------------------------------------------------------------------------------------%
\subsection{Input of the CESA}
\label{Subsec: Input of the CESA}

Since unweighed networks are considered in this study, the elements of adjacency matrix $A[N][N]$ of network $G$ with $N$ nodes and $E$ edges only consist of `1's and `0's, with the `1' indicating that an edge exists between node $i$ and node $j$ and the `0' representing no direct connection between them. So, the majority of the elements of the sparse network $G$ are `0's.

\begin{figure}[tbp]
\centerline{\epsfxsize=11cm \epsfbox{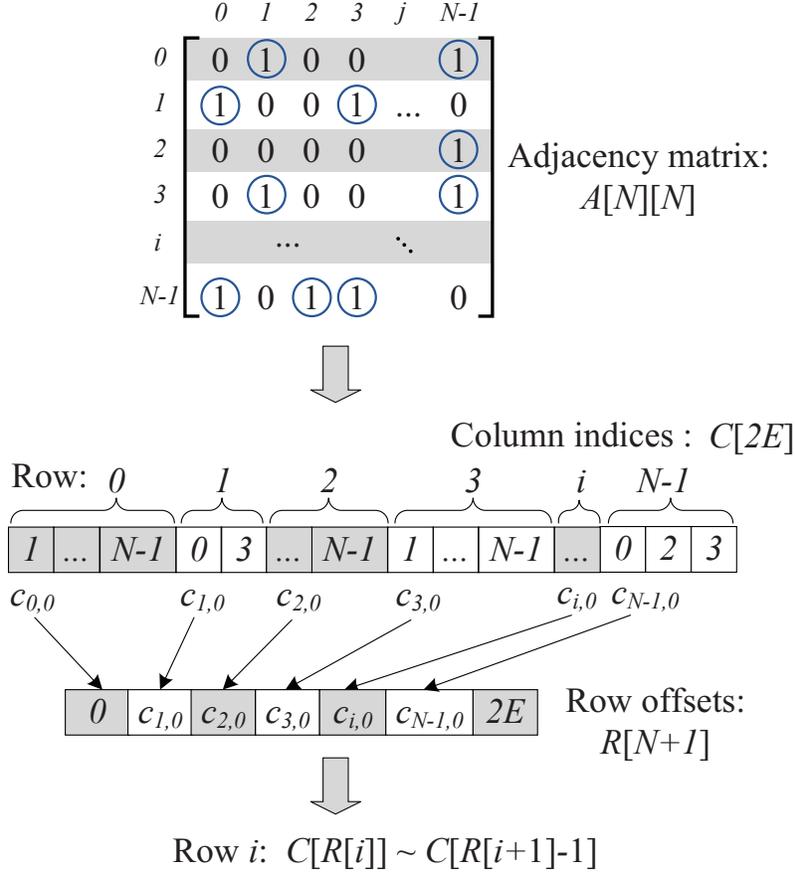}}
\caption{Obtaining $C[2E]$ and $R[N+1]$ from the adjacency matrix $A[N][N]$.}
\label{Fig: CSR}
\end{figure}

The CSR format of network $G$ consists of three one-dimensional arrays, namely, the column indices, the row offsets, and the values of non-zero elements. It can be easily converted from its adjacency matrix $A[N][N]$ or sparse adjacency matrix. As described above, the value of non-zero elements in this study is `1'. Thus, the CSR format used here is composed of two arrays: the array of column indices $C[2E]$ to store the column indices $j$ of the `1's in the adjacency matrix, and the array of row offsets $R[N+1]$ to store the starting offset of a new row $c_{i,0}$ in $C[2E]$. Using two arrays instead of three arrays further reduces the space requirement. Fig. \ref{Fig: CSR} illustrates how to obtain $C[2E]$ and $R[N+1]$ of network $G$ from its adjacency matrix $A[N][N]$. The last element of $R[N+1]$ is $2E$, i.e., twice of the number of edges in the network $G$. In this way, it is easy to extract the direct neighbors of node $i$, which are $C[R[i]]$ to $C[R[i+1]-1]$. Another important feature of the CSR format is that the elements of each row in the column indices can be out of order. Therefore, $C[2E]$ and $R[N+1]$ of network $G$ can also be converted from its unordered sparse adjacency matrix $A_s[2E][2]$, further saving the sorting time. Here, the sparse adjacency matrix $A_s[2E][2]$ is composed of the positions of `1's in the adjacency matrix $A[N][N]$, including the row indices $i$ and the column indices $j$. Because the sparse adjacency matrix $A_s[2E][2]$ usually requires less memory than the adjacency matrix $A[N][N]$ for a sparse network, we get $C[2E]$ and $R[N+1]$ of the network from its sparse adjacency matrix $A_s[2E][2]$ in our practical calculations.

As a result, the computational complexity to obtain the input (i.e., the $C[2E]$ and $R[N+1]$) of the CESA is significantly reduced. The time complexity is reduced from $O(N^3)$ to $O(E)$ and the space complexity is reduced from $O(N^2)$ to $O(E+N)$. For example, for an unweighted network $G$ with $100,000$ nodes and $1,000,000$ edges, the required memory for storing $C[2E]$ and $R[N+1]$ is as low as a few megabytes. Our experimental tests show that the actual memory requirement is approximately $8.0$ MB for $C[2E]$ and $R[N+1]$. This is compared to $37.3$ GB for the shortest-path distance matrix.

%-------------------------------------------------------------------------------------------------------------------%
\subsection{Algorithm design}
\label{Subsec: Algorithm design}

\begin{figure}[tbp]
\centerline{\epsfxsize=11cm \epsfbox{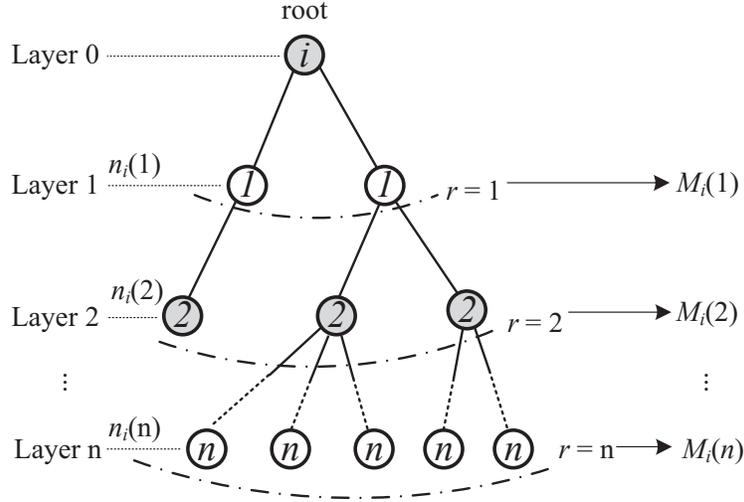}}
\caption{Obtaining the $M_i(r)$ with the BFS algorithm for the center node $i \in S_c$.}
\label{Fig: BFS}
\end{figure}

In the CESA, we employ the BFS algorithm to obtain the $M_i(r)$ as required in Eq. (\ref{Eq: MFA-LR}). Fig. \ref{Fig: BFS} shows the BFS process of searching the neighbor nodes of each layer of center node $i$. More specifically, the BFS algorithm is applied with the center node $i$ as the root. All other nodes of network $G$ are divided into different layers. Then, the shortest-path distance between center node $i$ and the other nodes can be easily obtained. As seen from Fig. \ref{Fig: BFS}, the number of nodes in the $l$th layer, $n_i(l)$, matches with the number of nodes whose distance from the center node $i$ equals to $r=l$. In this way, the number of nodes within the radius $r$ of the center node $i$ can be calculated by accumulating the number of nodes in each layer as shown in Eq. (\ref{Eq: Mir}).
\begin{equation}\label{Eq: Mir}
  M_i(r) = \sum n_i(l), ~~\forall l \leq r, \forall r\leq d.
\end{equation}

Now, the CESA algorithm is summarized as follows:
\begin{enumerate}
\item[(i)] Obtain the $C[2E]$ and $R[N+1]$ of network $G$;

\item[(ii)] A number of nodes are randomly selected as elements of the center-node set $S_{c}(r)$;

\item[(iii)] Set the node $i~(i \in S_{c}(r))$ as the center node of the sandbox;

\item[(iv)] Based on the inputs $C[2E]$ and $R[N+1]$, conduct the BFS with the center node $i$ as the root and then calculate the $M_{i}(r)$ through the Eq. (\ref{Eq: Mir});

\item[(v)] For each $q$, calculate the $[M_{i}(r)]^{q-1}$;

\item[(vi)] For all different center node $i$, repeat steps (iv) to (v) to calculate the $[M_{i}(r)]^{q-1}$;

\item[(vii)] Calculate the statistical averages $\langle [M_{i}(r)]^{q-1} \rangle$ of $[M_{i}(r)]^{q-1}$ over all sandboxes of radius $r$ and then use them to calculate the $\tau_{q}$ or $D_{q}$ with a linear regression according to Eq. (\ref{Eq: MFA-LR}).
\end{enumerate}

For the second step of the CESA, a number of nodes are randomly selected as elements of the center-node set $S_c$, of which the size $N_c$ is proportional to the network size $N$. The center-node set $S_c$ remains the same in the following steps once it is determined. This reduces the complexity of the existing sandbox algorithm. Since the statistical average values $\langle[M_i(r)]^{q-1}\rangle$ are used for linear regressions as shown in Eq. (\ref{Eq: MFA-LR}), this change does not impact the final multifractal results from the statistical perspective. This can be verified by our experiments in Section \ref{Sec: Experiment}.

In addition, with the shortest-path distance matrix in the existing sandbox algorithm, the network diameter $d$ in Eq. (\ref{Eq: MFA-LR}) can be obtained straightaway. However, this is not directly available for the CESA with the CSR format as the input. At first glance, this parameter $d$ is necessary for analyzing the multifractal results as required in Eq. (\ref{Eq: MFA-LR}). In fact, this parameter does not affect the MFA results because it is fixed for a given network $G$. Furthermore, it is seen from the Eq. (\ref{Eq: MFA-Dq}) that the range of radius $r$ used for linear regressions should be selected in the small-scale range of $r$. Of course, we can also use the observed network diameter $d'$ as an approximation of the actual network diameter $d$, which is updated to the maximum depth of the trees obtained by the BFS rooted by center node $i \in S_c$ in step (iv). As a result, whether $d'$ equals to $d$ or not depends on whether one of the nodes with the longest distance in network $G$ is selected as a center node in $S_c$. This can be certainly guaranteed when the size of center-node set $S_c$, $N_c$, equals to the network size $N$. Or it can be guaranteed by adding the node with longest distance into the center-node set $S_c$. Actually, our experiments to be presented later in Section \ref{Sec: Experiment} show that $d'$ is almost equal to $d$ when $N_{c}$ equals to $10\%$ of the network size $N$.

%-------------------------------------------------------------------------------------------------------------------%
\subsection{Complexity analysis of CESA}
\label{CESA: Complexity analysis}

As mentioned above, the time complexity of getting the input of the CESA is $O(E)$. The BFS is run for each of the center nodes in the set $S_c$, implying that it executes $N_c$ times altogether. Since the time complexity of the BFS algorithm is known to be $O(N+E)$, the overall time complexity of the CESA can be expressed as $O(E+N_c(N+E+N_q))$. For large-scale complex networks, $N_q$ is negligible compared to $(N+E)$. Thus, the overall time complexity of the CESA can be further expressed as $O(N_c(N+E))$. As $N_c$ is usually smaller than $N$ and $E$ is much smaller than $N^2$, the time complexity $O(N_c(N+E))$ of the CESA is much smaller than the time complexity $O(N^3)$ of the existing sandbox algorithm. For the extreme case when all the nodes are selected as center nodes ($N_c=N$), the overall time complexity of the CESA would be $O(N(N+E))$, which is still much less than $O(N^3)$ of the existing sandbox algorithm.

The overall space complexity of our CESA is $O(N+E)$, which is mainly determined by the size of the CSR format, i.e., the $C[2E]$ and $R[N+1]$. For many complex networks, it is much smaller than the space complexity $O(N^2)$ of the existing sandbox algorithm. This can be verified from many real-world networks that are sparse in nature.

It is known that a network of size $N$ has at most $N(N-1)/2$ undirected edges. In this case, all nodes in the network are pair-wise connected by an edge. Therefore, if the number of edges of network $G$ we considered is approximately $O(N^{2})$, the overall time and space complexities of the CESA would be $O(N_{c}N^2)$ and $O(N^2)$, respectively. Further, when combined with the $N_c = N$ mentioned above, the overall time and space complexities are going to be $O(N^3)$ and $O(N^2)$, respectively. This is the worst-case, which makes the CESA consuming the same time and space complexities as the existing sandbox algorithm. This is to say that even in the worst-case, our CESA is still no worse than the existing sandbox algorithm. However, since most model and real-world networks are sparse networks, our CESA generally has great advantages in time and space costs.

In summary, through the redesign of the computational process and using the two arrays of the CSR format of the adjacency matrix as the input, the time complexity is reduced from $O(N^3)$ to $O(N_c(N+E))$ and the space complexity is reduced from $O(N^2)$ to $O(N+E)$. These make the CESA more efficient in both time and space for calculating the mass exponents $\tau_{q}$ or the generalized fractal dimensions $D_{q}$ of large-scale complex networks.

%%%%%%%%%%%%%%%%%%%%%%%%%%%%%%%%%%====================================================%%%%%%%%%%%%%%%%%%%%%%%%%%%%%%%
\section{Experimental studies}
\label{Sec: Experiment}

This section conducts experiments to verify the effectiveness and efficiency of our CESA for MFA of large-scale complex networks. We use the ($u$,$v$)-flower model networks as verification networks. Both the computational accuracy and complexity of the proposed CESA are investigated. Then, the CESA is applied to a few real-world complex networks of large scale. For all these experiments, the number of selected center nodes $N_c$ equals to $10\%$ of the network size $N$, i.e., $N_c=0.1N$.

%-------------------------------------------------------------------------------------------------------------------%
\subsection{Algorithm verification}

\begin{figure}[tbp]
\centerline{\epsfxsize=11cm \epsfbox{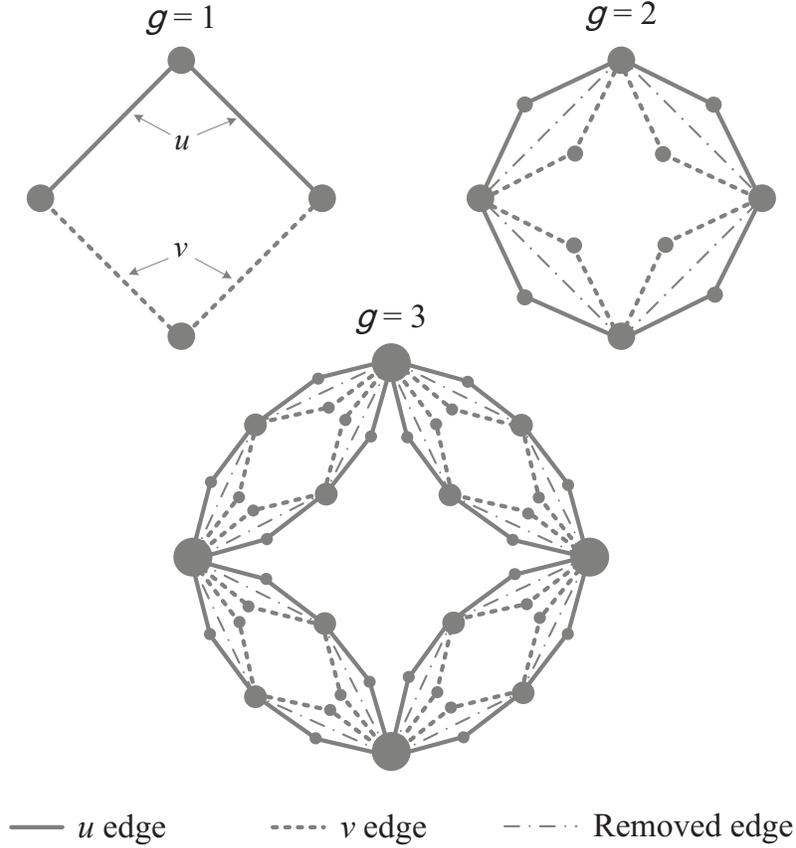}}
\caption{The 1st, 2nd, and 3rd generation ($u$,$v$)-flower network with $u$=2 and $v$=2.}
\label{Fig: UV-Flower}
\end{figure}

In 2007, Rozenfeld \textit{et al.} proposed the ($u$,$v$)-flower network model with the aim to understand the self-similarity and dimensionality of complex networks \cite{RHB07}. The network model is constructed recursively with known network scalability and deterministic multifractality. Thus, it has been used to verify the performance of some MFA algorithms \cite{FY11, LYA15, PM20}. In this study, we use the ($u$,$v$)-flower network model to generate verification networks of different scales for our experiments. These verification networks are generated recursively starting from a ring network with $u+v$ nodes and $u+v$ edges. Fig. \ref{Fig: UV-Flower} illustrates how the ($u$,$v$)-flower network with $u = 2$ and $v = 2$ is constructed recursively. From generation to generation, each edge in the previous generation is replaced by two parallel paths with length $u$ and $v$, respectively. In this way, the number of edges $E$ and the network size $N$ in each generation can be respectively calculated by
\begin{equation}\label{Eq: Edges}
  E = (u+v)^g,
\end{equation}
\begin{equation}\label{Eq: Nodes}
  N = (\frac{u+v-2}{u+v-1})(u+v)^g+\frac{u+v}{u+v-1}.
\end{equation}
As seen from Eqs. (\ref{Eq: Edges}) and (\ref{Eq: Nodes}), with the increase in the generation $g$, the number of edges $E$ increases exponentially, and the network size $N$ grows nearly exponentially. Therefore, the ($u$,$v$)-flower network model is quite suitable for generating large-scale networks in the verification of the feasibility of the proposed CESA. Another advantage of using ($u$,$v$)-flower networks as verification networks is that its mass exponents, $\tau_q$, can be theoretically determined as \cite{FY11, LYA15}
\begin{equation}\label{Eq: Tauq}
  \tau _q =
  \left \{
  \begin{array}{c}
    (q-1)\frac{\ln (u+v)}{\ln u},~q < \frac{\ln (u+v)}{\ln 2}, \\
    ~\\
     q \frac{\ln ((u+v)/2)}{\ln u},~q \geq \frac{\ln (u+v)}{\ln 2}.
  \end{array}
  \right.
\end{equation}
Thus, we can verify the effectiveness and accuracy of the proposed CESA by comparing the theoretical mass exponents $\tau_q$ in Eq. (\ref{Eq: Tauq}) with the numerical ones calculated from our CESA.

We first generate the 12th generation ($u$,$v$)-flower network with $u = 2$ and $v = 2$. The 12th generation ($u$,$v$)-flower network has $11,184,812$ nodes and $16,777,216$ edges. From our theoretical complexity analysis, the required memory space for the shortest-path distance matrix in the existing sandbox algorithm is in the order of a few hundreds of terabytes ($\approx 455.1$ TB). It is far beyond the computation capacity of the existing sandbox algorithm on normal desktop computers or even some high-performance computers. Now we use our CESA to perform the MFA for the network. According our theoretical analysis, the required memory space for the sparse data structure as input to the CESA is in the order of $100$ megabytes ($\approx 170.7$ MB).

\begin{figure}[tbp]
\centerline{\epsfxsize=11cm \epsfbox{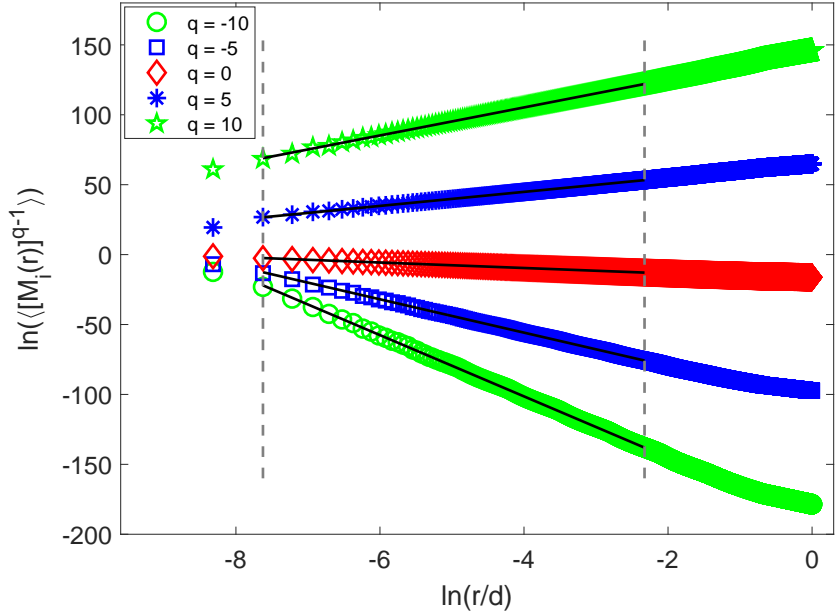}}
\caption{Solid lines (black lines) show the linear regressions for calculating the mass exponents $\tau_q$ of the 12th generation ($u$,$v$)-flower network with $u = 2$ and $v = 2$. The range between two dashed lines is used for linear regressions.}
\label{Fig: Scaling}
\end{figure}

\begin{figure}[tbp]
\centerline{\epsfxsize=11cm \epsfbox{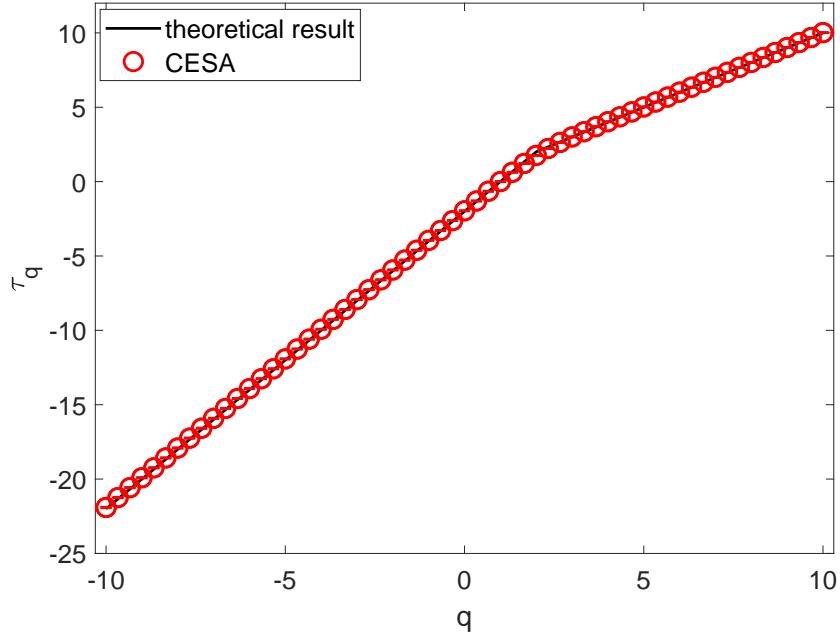}}
\caption{Mass exponents $\tau_q$ for the 12th generation ($u$,$v$)-flower network with $u = 2$ and $v = 2$. Solid line (black line) represents the mass exponent $\tau_{q}$ given by Eq. (\ref{Eq: Tauq}). The circles represent the numerical estimation of the $\tau_{q}$ calculated by the CESA. Each error bar takes twice the length of the standard deviation for all the results.}
\label{Fig: Tauq-UVFlower}
\end{figure}

From our CESA, Fig. \ref{Fig: Scaling} depicts the linear regressions of $\ln (\langle [M_i(r)]^{q-1} \rangle)$ versus $\ln (r/d)$, where the observed network diameter $d'$ is used as an approximation of $d$. These experimental results show good linearity. Here, we select [2,400] as the range of radius $r$ for linear regressions as shown in Fig. \ref{Fig: Scaling}. Then, we obtain the numerical mass exponents $\tau _q$ by the linear regressions in this linear rang based on Eq. (\ref{Eq: MFA-LR}). Fig. \ref{Fig: Tauq-UVFlower} illustrates the numerical and theoretical results of mass exponents $\tau _q$ with respect to $q$. It is seen from Fig. \ref{Fig: Tauq-UVFlower} that the mass exponents $\tau_q$ of the 12th generation ($u$,$v$)-flower network calculated by the CESA (circles) match well with the theoretical ones (solid line) obtained from the Eq. (\ref{Eq: Tauq}). This not only verifies the feasibility of our CESA on MFA of large-scale networks, but also experimentally demonstrates its accuracy and effectiveness. Although the computational process of our CESA and the existing sanbox algorithm are different, they share the same mathematical theory as introduced in Subsection \ref{Subsec: Sandbox}. This demonstrates that the proposed CESA in this study improves the computational efficiency without causing any sacrifice on the accuracy of the MFA results of complex networks. In addition, we also calculate the standard deviations of these mass exponents $\tau_{q}$. The calculated standard deviations are also shown as error bars in Fig. \ref{Fig: Tauq-UVFlower}, where each error bar takes twice the length of the standard deviation for all the results. However, these error bars are so short that they are almost invisible and become horizontal bars in the circles as shown in Fig. \ref{Fig: Tauq-UVFlower}, implying that our results of mass exponents $\tau_q$ calculated by the CESA are stable.

Next, we focus on the accuracy of the MFA results of the proposed CESA for ($u$,$v$)-flower networks with increasing network sizes. For this purpose, we generate ($u$,$v$)-flower networks with $u = 2$ and $v = 2$ from the 5th to the 12th generations, and apply the CESA to perform the MFA for these networks. In order to quantify the accuracy of the CESA for the MFA of these networks with different sizes, the relative standard error $E_{rs}$ \cite{LYA15}, the absolute square error $E_{as}$, and the absolute error $E_a$ are analyzed between theoretical and numerical values of mass exponents $\tau_q$. Let $\tau^t_q$ and $\tau^n_q$ denote the theoretical and numerical values of mass exponents $\tau_q$, respectively. Also, let $\bar{\tau^t}$ denote the average of $\tau^t_q$. Then, errors $E_{rs}$, $E_{as}$, and $E_a$ are respectively defined by
\begin{equation}\label{Eq: Ers}
  E_{rs} = \frac{\sqrt{\frac{1}{N_q}\sum\limits_{q}(\tau^t_q-\tau^n_q)^2}}{\sqrt{\frac{1}{N_q}\sum\limits_{q}(\tau^t_q-\bar{\tau^t})^2}},
\end{equation}

\begin{equation}\label{Eq: Eas}
  E_{as} = \sum\limits_{q}(\tau^t_q-\tau^n_q)^2,
\end{equation}

\begin{equation}\label{Eq: Ea}
  E_a = \sum\limits_{q}|\tau^t_q-\tau^n_q|.
\end{equation}

\begin{figure}[tbp]
\centerline{\epsfxsize=11cm \epsfbox{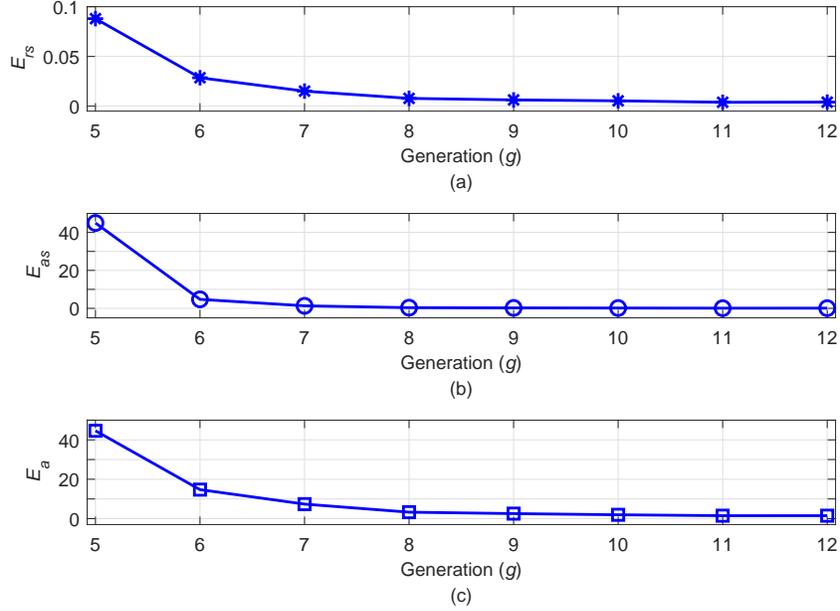}}
\caption{Error analysis for the mass exponents $\tau _q$ of the ($u$,$v$)-flower network with $u=2$ and $v=2$ from $684$ nodes of the 5th generation to $11,184,812$ nodes of the 12th generation: (a) relative standard error $E_{rs}$, (b) absolute square error $E_{as}$, (c) absolute error $E_{a}$.}
\label{Fig: Error}
\end{figure}

Fig. \ref{Fig: Error} depicts these errors between the numerical mass exponents $\tau_q$ of CESA and theoretical ones of ($u,v$)-flower network as the network size $N$ increases. The experimental errors of CESA, including $E_{rs}$, $E_{as}$, and $E_a$, decrease significantly as the network size $N$ increases from $684$ nodes of the 5th generation to $11,184,812$ nodes of the 12th generation. This indicates that the CESA improves the accuracy of the MFA results with the increase in the network size $N$. Therefore, calculating large-scale networks for MFA is beneficial.

We take the execution time of the CESA as an indication of its time complexity. On a normal desktop computer with a 4-core CPU and 16 GB memory, the execution time performance for computing the mass exponents $\tau_{q}$ of ($u$,$v$)-flower networks with $u=2$ and $v=2$ from the 7th (with $10,924$ nodes and $16,384$ edges) to the 12th generation (with $11,184,812$ nodes and $16,777,216$ edges) is recorded. The experimental results of the execution time are compared with the theoretical time complexity of the CESA ($O(N_c(N+E))$) and the existing sandbox algorithm ($O(N^3)$).

\begin{figure}[tbp]
\centerline{\epsfxsize=11cm \epsfbox{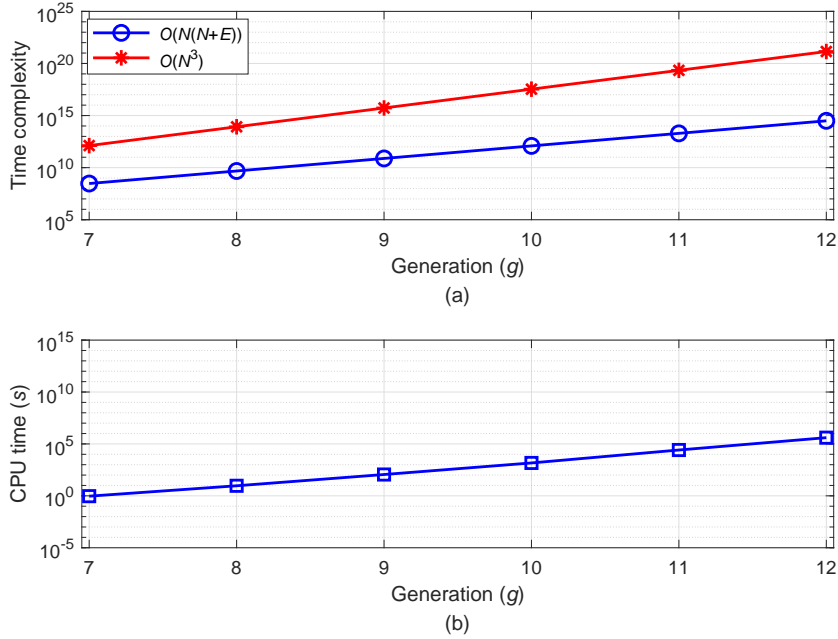}}
\caption{(a) Time complexity of the CESA and the existing sandbox algorithm when $N_c = N$. (b) CPU time of the CESA when $N_c = N/10$.}
\label{Fig: TC-CPU}
\end{figure}

Fig. \ref{Fig: TC-CPU}(a) compares the theoretical time complexity of the CESA and the existing sandbox algorithm, where the $N_c = N$ is considered. Here, semi-log graph is used since the network size, $N$, and the number of edges, $E$, increase significantly from generation to generation as shown in Eqs. (\ref{Eq: Edges}) and (\ref{Eq: Nodes}). As a result, the time complexity of the CESA, $O(N_c(N+E))$, and that of the existing sandbox algorithm, $O(N^3)$, both increase nearly exponentially. Even in the extreme case of $N_c = N$, however, it is observed that $O(N^3)$ has greater slope than $O(N_c(N+E))$, indicating that the computational burden of the existing algorithm increases much faster than that of the CESA.

Fig. \ref{Fig: TC-CPU}(b) shows the CPU time of the CESA on our desktop computer when 10\% of the nodes are selected as center nodes (i.e., $N_c=0.1N$). For the 7th generation ($u$,$v$)-flower network, the execution time of the CESA is about $1$ second when the network size and the number of edges are both in the order of ten thousands. It increases to $111$ hours for the 12th generation when both values increase to more than ten millions. With an effort of $111$ hours, our CESA gives the computing results. In contrast, the existing sandbox algorithm fails to handle this network completely. Overall, the slope of the CPU time of our CESA is more close to that of the $O(N_c(N+E))$ as shown in Fig. \ref{Fig: TC-CPU}, which experimentally verifies that the time complexity of the CESA is reduced to quadratic.

In addition, it is worth mentioning that although the desktop computer used for these experiments is equipped with a 4-core CPU and 16 GB memory, the real-time monitoring of the computing process shows that only 13\% of the CPU resource and 10\% of the memory are actually used by the CESA for MFA of the 12th generation ($u$,$v$)-flower network. Therefore, the network size that can be analyzed with the proposed CESA on a normal desktop computer can be much greater, enabling MFA for complex networks of a larger scale.

%-------------------------------------------------------------------------------------------------------------------%
\subsection{Applications}

Finally, we apply our CESA to conduct the MFA for a few real-world complex networks of large scale. Provided on Stanford Large Network Dataset Collection \cite{SLNDC}, these complex networks include:
\begin{enumerate}
	\item[1)] the LiveJournal social network with $3,997,962$ nodes and $34,681,189$ edges;
	\item[2)] Orkut social network with $3,072,441$ nodes and $117,185,083$ edges;
	\item[3)] Youtube social network with $1,134,890$ nodes and $2,987,624$ edges;
	\item[4)] California road network with $1,965,206$ nodes and $2,766,607$ edges;
	\item[5)] Pennsylvania road network with $1,088,092$ nodes and $1,541,898$ edges;
	\item[6)] Texas road network with $1,379,917$ nodes and $1,921,660$ edges; and
	\item[7)] the autonomous systems graph by Skitter with $1,696,415$ nodes and $11,095,298$ edges.
\end{enumerate}
For these large-scale networks, the existing sandbox algorithm fails to give MFA results on a normal desktop computer due to the high complexity of the algorithm. In comparison, our CESA works well on a normal desktop computer for all these networks due to the much reduced algorithm complexity.

\begin{figure}[tbp]
\centerline{\epsfxsize=11cm \epsfbox{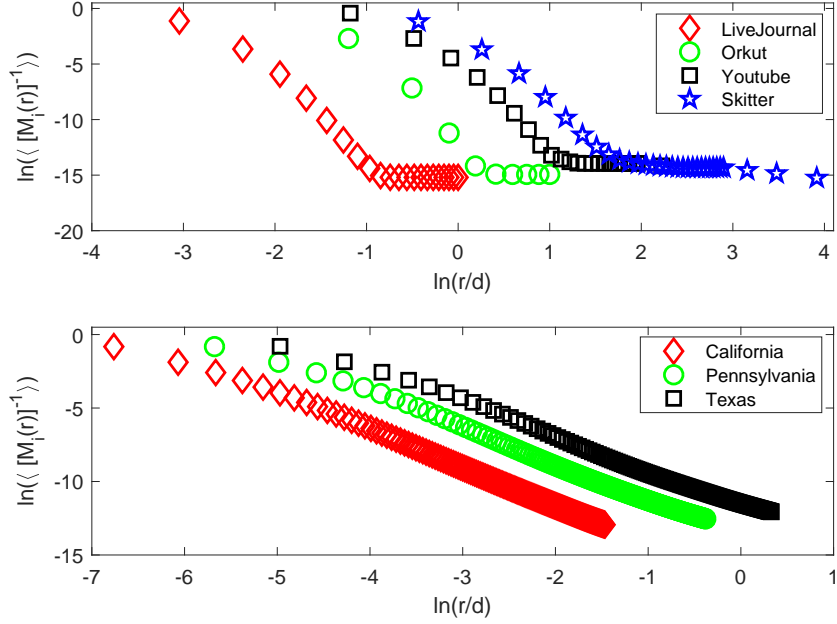}}
\caption{MFA results of some real-world and large-scale networks for $q = 0$ by using the CESA.}
\label{Fig: App}
\end{figure}

Fig. \ref{Fig: App} shows the MFA results of our CESA for these complex networks under $q = 0$. It is observed from this figure that the road networks show an apparent power-law behavior for $q=0$, indicating their fractal characteristics. However, there is lack of good power-law relation in these networks for some non-zero values of $q ~(q\neq0)$. In our understanding, this phenomenon is not uncommon in many real-world networks. However, it is still not clear why these networks do not possess the clear multifractality like some model networks, e.g., the ($u$,$v$)-flower model network. To answer this question, deeper investigations into various real-world and model networks need to be conducted. It is also seen from Fig. \ref{Fig: App} that the investigated social networks, namely, LiveJournal, Orkut, Youtube, and Skitter, do not have the clear fractality. They look more like a shifted power-law (i.e., modified power-law or Mandelbrot's law) behaviour or a pure exponential decay as mentioned in Ref. \cite{SHM05}. This is not surprising because Song \textit{et~al.} have pointed out that the lack of clear fractality in some networks might be due to incomplete information of these networks \cite{SHM05}.

In summary, with its much reduced complexity over existing sandbox algorithm, our CESA presented in this paper enables us to reveal the fractality and multifractality of these and other real-world complex networks of large scale.

%%%%%%%%%%%%%%%%%%%%%%%%%%%%%%%%%%====================================================%%%%%%%%%%%%%%%%%%%%%%%%%%%%%%%
\section{Conclusions}
\label{Sec: Conclusions}

A computationally-efficient sandbox algorithm (CESA) has been presented in this paper for MFA of large-scale complex networks. Distinct from the existing sandbox algorithm that uses the shortest-path distance matrix to obtain the required information for MFA of complex networks, our CESA applies the BFS to directly search the neighbor nodes of each layer of center nodes, and then to retrieve the required information. Our CESA's input is a sparse data structure derived from the CSR format for compressed storage of the adjacency matrix of large-scale networks. As a result, the complexity is significantly reduced. For a complex network with $N$ nodes, $E$ edges, and $N_c$ center nodes, the time complexity is reduced from $O(N^3)$ to $O(N_c(N+E))$, and the space complexity reduced from $O(N^2)$ to $O(N+E)$. $N_c$ is usually smaller than $N$. As most model and real-world networks are sparse, $E$ is much smaller than $N^2$. Therefore, the reduction of both time complexity and space complexity from our CESA is significant over the existing sandbox algorithm. Experiments of our CESA have been conducted for the MFA of some model and real-world large-scale networks. The MFA results of ($u$,$v$)-flower model networks from the 5th to the 12th generations verify that the CESA is not only effective but also computationally efficient and feasible. More specifically, the presented CESA has been successfully applied to the MFA on a normal desktop computer for the 12th generation ($u$,$v$)-flower network with $11,184,812$ nodes and $16,777,216$ edges. Such a scale of complex networks is far beyond the limit of the existing sandbox algorithm on the same desktop computer. A further analysis on CPU time and mass exponents has shown that the CESA reduces the time complexity to quadratic without sacrificing the accuracy of the MFA results. Moreover, we have also found that the accuracy of the MFA results can be improved significantly with the increase in the size of a theoretical network, further verifying the value of this study on a computationally-efficient algorithm for the MFA of large-scale complex networks. Finally, the proposed CESA has been applied to a few real-world complex networks of large scale. The clear fractality has been observed for large-scale road networks of some cities.

%%%%%%%%%%%%%%%%%%%%%%%%%%%%%%%%%%====================================================%%%%%%%%%%%%%%%%%%%%%%%%%%%%%%%
\section*{Acknowledgement}
This project was supported in part by the Australian Research Council (ARC) through the Discovery Project Scheme (Grant No. DP170103305), in part by the National Natural Science Foundation of China (Grant Nos. 61702369, 11801483, and 11871061), and in part by the Natural Science Foundation of Hunan Province of China (Grant No. 2019JJ50575).

%%%%%%%%%%%%%%%%%%%%%%%%%%%%%%%%%%====================================================%%%%%%%%%%%%%%%%%%%%%%%%%%%%%%%
\setstretch{1.0}

\end{document}